\documentstyle[11pt,paspconf,epsf,psfig]{article}

\begin{document}

\title{The ESO Nearby Abell Cluster Survey: \protect\\
Kinematics of Galaxies in Clusters}

\author{Andrea Biviano}
\affil{Osservatorio Astronomico di Trieste, I}

\author{Alain Mazure, Christophe Adami\altaffilmark{1}}
\affil{IGRAP, Laboratoire d'Astronomie Spatiale, Marseille, F}

\author{Peter Katgert, 
Roland den Hartog\altaffilmark{2},
Pascal de Theije}
\affil{Sterrewacht Leiden, NL}

\author{George Rhee}
\affil{University of Nevada, Las Vegas, USA}

\altaffiltext{1}{Dearborne Observatory, NWU, Evanston IL, USA}
\altaffiltext{2}{ESA, Astroph. Div., Space Science Dept., ESTEC, Noordwijk, NL}

\begin{abstract}
We summarize several results based on the velocity data-set for
cluster galaxies provided by ENACS (the ESO Nearby Abell Cluster
Survey). A more general description of ENACS is given in the companion
review by Katgert et al., in this volume.
 
We describe the distribution of velocity dispersions of a complete
sample of rich galaxy clusters, and compare it to the distribution of
cluster X-ray temperatures, and with predictions of theoretical
models.
 
We then address the issue of the existence of a Fundamental Plane (FP)
for rich clusters, first suggested by Schaeffer et al.  We confirm the
existence of this FP with the ENACS data-set. The cluster FP is
different from the FP of elliptical galaxies, and from the virial
prediction. Some implications of the cluster FP are discussed.
 
Finally, we describe the phase-space distributions of different
populations of cluster galaxies. Different galaxy classes are defined
according to their morphological or spectral type, and the presence of
emission-lines. Star-forming (late-type) galaxies have a velocity
dispersion profile that (in combination with their wider spatial
distribution) is suggestive of first infall into the cluster. On the
contrary, quiescent (early-type) galaxies show evidence of a
dynamically relaxed distribution.

\end{abstract}


\section{Introduction}
The ESO Nearby Abell Cluster Survey (ENACS, see Katgert et al. 1996,
1998) is currently the largest homogeneous dataset of redshifts for galaxies in
clusters. At its completion, in 1994, it almost doubled the number of
available redshifts for cluster galaxies. The survey was designed in
such a way as to provide (in combination with the literature) a
complete, volume-limited sample of $\geq 100$ rich Abell clusters,
with robust estimates of their kinematical parameters, such as, in
particular, the velocity dispersion. Previous to ENACS, most results
on the cluster properties, in particular on those concerning the
kinematics, were based either on a few clusters only (e.g. Colless \&
Hewett 1987, Dressler \& Shectman 1988), or on heterogeneous
collections of cluster data from many sources (e.g. Girardi et
al. 1993). ENACS has allowed for the first time to analyse the
kinematics of a large homogeneous cluster data-set, where possible
biases and selection effects were well under control.

In this review, we summarize the main results obtained so far with the
ENACS data-set, on the kinematical properties of clusters.
In \S~2 we present the distribution of velocity dispersions
of a complete volume-limited cluster sample (Mazure et al. 1996). In
\S~3 we describe the existence and characteristics of the Fundamental
Plane of galaxy clusters in the luminosity, radius, velocity
dispersion space (Adami et al. 1998a). Finally, in \S~4 we deal with
the phase-space distributions of different populations of cluster
galaxies (Biviano et al. 1997, Adami et al. 1998b, de Theije \&
Katgert 1998).

We refer the reader to the companion review of Katgert et al. (these
proceedings) for a thorough description of the ENACS data-set, and for
a review of those results which are not based on the velocity data-set.

\section{The Distribution of Cluster Velocity Dispersions}
The distribution of cluster masses can be used to constrain theories
of cluster formation and evolution (see, e.g., Bahcall \& Cen 1992).
Since mass estimates for clusters are difficult to obtain, and often
require a-priori assumptions on the dynamical state of the clusters
(see, e.g., Biviano et al. 1993), mass gauges, such as the velocity
dispersion and the intra-cluster gas temperature, have often been used
instead. Before ENACS, there have been several attempts to derive the
distribution of cluster velocity dispersions ($\sigma_v$'s), but these
were based either on heterogeneous data-sets (Frenk et al. 1990,
Girardi et al. 1993), or on homogeneous but small data-sets (Zabludoff
et al. 1993).

\subsection{Completeness of The Sample, and The Density of Rich Clusters}
By combining the ENACS data-set with 1000 redshifts for 37 clusters
drawn from the literature, we built a volume-limited sample of 128
Abell clusters with richness $R \geq 1$, in a 2.55 sr cone, within a
redshift limit $z \leq 0.1$. The data-set is overall very homogeneous,
since most of the data come from ENACS. For 80 of the 128 clusters in
the complete sample, we obtained a reliable estimate of the
$\sigma_v$, based on at least 10 (but very often between 30 and 150)
redshifts for cluster members.

A critical point we addressed in our analysis is that of
completeness. Since our sample is selected from the catalogues of
Abell (1958) and Abell et al. (1989), we had to apply a correction for
the incompleteness of these catalogues. For this purpose, we used the
Edinburgh-Durham Cluster Catalogue (Lumsden et al. 1992) as a
comparison sample. The resulting {\em corrected} space density of
richness $R \geq 1$ clusters is: 
\begin{displaymath}
8.6 \pm 0.6 \times 10^{-6} h^3 Mpc^{-3}
\end{displaymath}
slightly higher than most previous estimates, but significantly lower
than that obtained by Scaramella et al. (1991). This density estimate
has recently been used by Carlberg et al. (1997) to constrain the
redshift evolution of galaxy cluster densities.

\subsection{The $\sigma_v$-distribution}
Completeness to a given overdensity implies completeness to a given
$\sigma_v$, through the relation between $\sigma_v$'s and richness
counts. However, the large spread in this relation implies that a
sample complete to a given richness is biased against low-$\sigma_v$
clusters. As a consequence, our sample, complete for richness $R \geq
1$, is only complete for $\sigma_v \geq 800$~km/s.

\begin{figure}
\centerline{\psfig{file=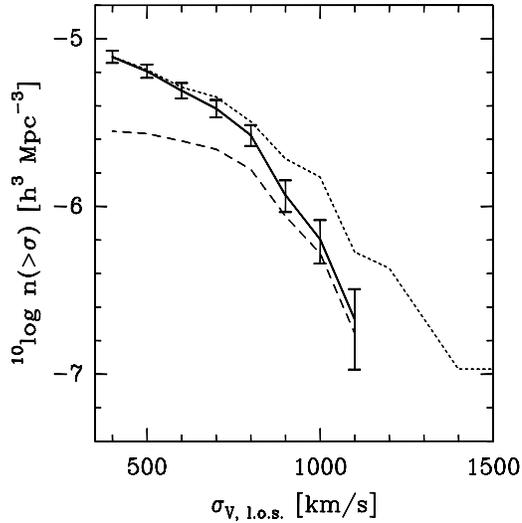,height=7cm,width=7cm}}
\caption{The cumulative distribution of cluster velocity
dispersions. Solid line: for the sample of clusters complete in 2D richness. 
Dotted line: for the sample of clusters complete in 2D richness, {\em before
interloper removal.} Dashed line: for the sample of clusters complete in 3D
richness.}
\label{f-sigmaint}
\end{figure}

As an alternative of using 2D richness counts, we also estimated
3D richnesses, which we obtained from the total galaxy counts,
corrected for the fraction of background and foreground galaxies, as
estimated from ENACS. The final $\sigma_v$ distributions we obtained
for the 2D-richness complete sample and the 3D one, were consistent.
This can be seen in Figure~\ref{f-sigmaint}, where the two
distributions are plotted. Note that the 3D-richness sample is only
complete down to $\sigma_v = 900$~km/s, slightly higher than the
completeness limit of the 2D-richness complete sample.

Velocity dispersions were computed by applying the interloper
removal procedure of den Hartog \& Katgert (1996). In
Figure~\ref{f-sigmaint} we show for comparison the distribution one
would obtain by estimating $\sigma_v$ {\em without} interloper removal
(dotted line). The removal of interlopers is clearly essential in
properly estimating the $\sigma_v$-distribution.

\begin{figure}
\centerline{\psfig{file=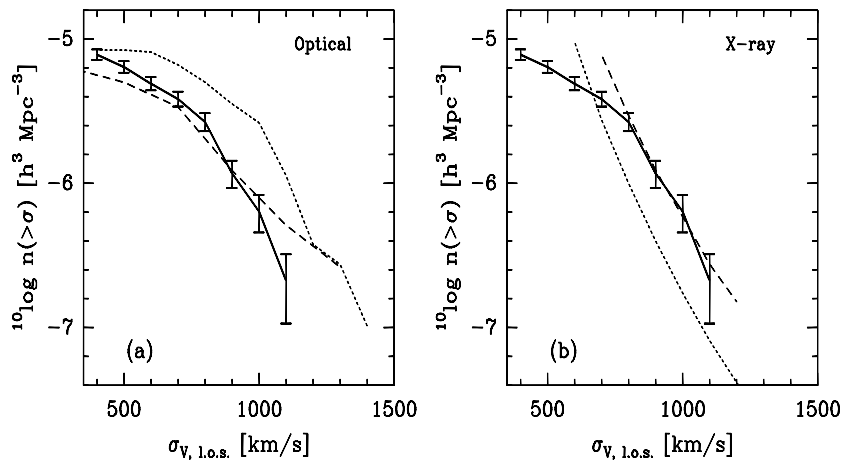,height=6cm,width=12cm,angle=270}}
\caption{Comparison of our $\sigma_v$-distribution (solid line) with
distributions from the literature. Panel a): optical data (dotted
line, Girardi et al. 1993; dashed line, Zabludoff et al. 1993); panel
b): X-ray data (dotted line: Edge et al. 1990; dashed line, Henry \&
Arnaud 1991).}
\label{f-sigmacomp}
\end{figure}

Our distribution is in good agreement with the distribution obtained
by Zabludoff et al. (1993) only for $\sigma_v \leq 900$~km/s, but for
larger values of $\sigma_v$ they obtained a significantly flatter
slope, possibly due to an incorrect removal of interlopers (see
Figure~\ref{f-sigmacomp}). The distribution of Girardi et al. (1993),
on the other hand, is systematically higher all along the $\sigma_v$
range.

Using the relation $\beta = \sigma_v^2/(k T_X /\mu m_H) = 1$, we found
an excellent agreement between our $\sigma_v$ distribution and the
distribution of cluster X-ray temperatures ($T_X$) of Henry \& Arnaud
(1991).  Note that the other $T_X$ distribution, that of Edge et al.
(1990), had a wrong normalisation, according to Ebeling et al. (1997).
The good agreement found for a value $\beta=1$, suggests that $T_X$
and $\sigma_v$ are both valid measures of the cluster gravitational
potential.

Finally we used our distribution of $\sigma_v$'s to constrain the
standard Cold Dark Matter models of Frenk et al. (1990) and of van
Kampen (1994). Even if there is not a perfect agreement between these
models , in general the comparison of both models with the observed
distribution indicates that a large bias parameter, $b \geq 2$, is
required. For the commonly accepted low values of the bias parameter,
the standard CDM model clearly overpredicts the number of
high-$\sigma_v$ clusters. The observed relative proportions of high-
and low-$\sigma_v$ clusters is also poorly reproduced in the models.

\section{The Fundamental Plane of clusters of galaxies}
It is well known that elliptical galaxies populate a fundamental plane
(FP, hereafter) in the (radius, internal velocity dispersion,
luminosity)-space (e.g. Dressler et al. 1987). Here we are interested
in the existence of a FP {\em for clusters of galaxies.} The existence
of a galaxy cluster FP was claimed by Schaeffer et al. (1993). Their
claim was based on a heterogeneous data-set of 16 galaxy clusters, so
we deemed it interesting to re-address this issue by using a larger
and more homogenous data-set.

\subsection{The Sample; Luminosities, Radii and Velocity Dispersions}
We selected 29 ENACS clusters, each with at least 10 galaxy redshifts,
for which magnitudes and positions from the Cosmos catalogue
(Heydon-Dumbleton et al. 1989, and MacGillivray, private communication)
were available to us. The Cosmos data were used to determine the
characteristic scales, $R$, and luminosities, $L$, of the clusters,
while the ENACS data provided the $\sigma_v$-estimates.

Four different kinds of number density profiles were fitted to the
Cosmos data in order to derive $R$, i.e.: King, Hubble,
de~Vaucouleurs, and a 2D analogue of the profile of Navarro et
al. (see Adami et al. 1998a and references therein). Best fits were
obtained for the King and Hubble profiles (Adami et al. 1998c; Katgert
et al., these proceedings).

Luminosities were determined from the Cosmos $b_j$ galaxy magnitudes,
which were K-corrected, corrected for the galactic absorption, and
transformed to absolute magnitudes. The total luminosity of all Cosmos
galaxies in a given cluster field was then corrected for the
contribution of non-members, as estimated on the basis of the
available (ENACS) redshifts. Extrapolation beyond the completeness
magnitude of the Cosmos catalogue, was done using a Schechter (1976)
luminosity function. In a few cases, bright ENACS galaxies were found
with no Cosmos counterpart; we added the luminosities of these bright
galaxies to the total cluster luminosity.

The cluster $\sigma_v$'s were computed from the ENACS data, using the
biweight estimator (Beers et al. 1990), and the interloper removal
procedure of den~Hartog \& Katgert (1996).

Finally, we defined and evaluated the 'contrast' parameter, defined as
the overdensity of the galaxy number counts in the cluster region,
relative to the local background.

\subsection{The FP}
\begin{figure}
\centerline{\psfig{file=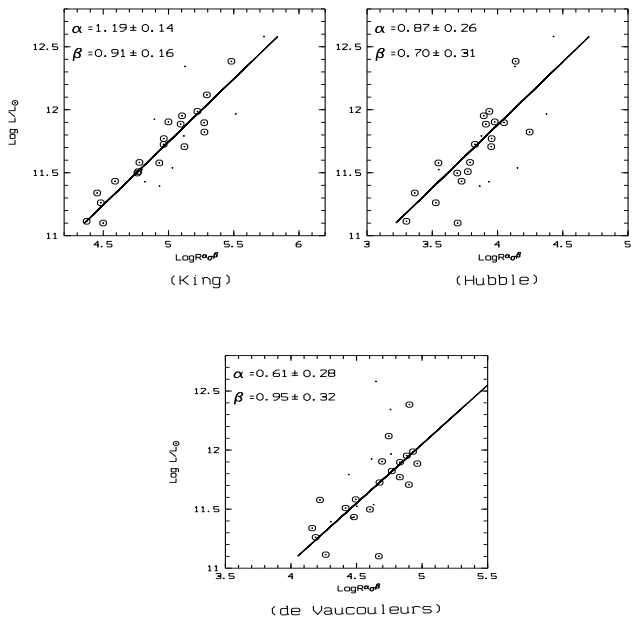,height=9cm,width=9cm,angle=180}}
\caption{The $L$-$R$-$\sigma_v$ relation for the King, Hubble and
de~Vaucouleurs profiles. This is an edge-on view of the FP defined by
our sample of galaxy clusters.  Dotted circles represent clusters of
highest contrast, while dots represent the less contrasted ones.}
\label{f-fpkh}
\end{figure}

\begin{figure}
\centerline{\psfig{file=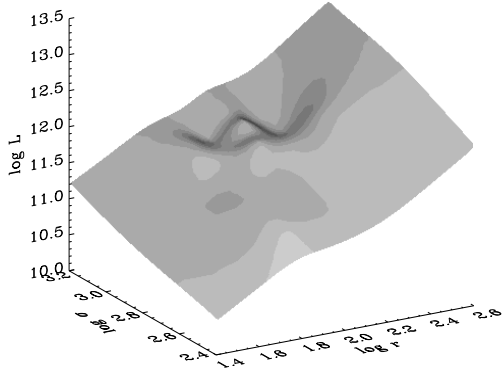,height=6cm,width=8cm}}
\caption{The FP defined by our sample of galaxy clusters in 3D.  The
surface shown is the minimum curvature spline surface fitted to the
($L$, $R$, $\sigma_v$) data. $R$ here is the King core-radius.}
\label{f-fp3d}
\end{figure}

We looked whether the $R$, $L$ and $\sigma_v$ parameters are related
in our cluster sample. No relation was found between $\sigma_v$ and
$R$. On the other hand, $R$ and $L$ {\em are} correlated, as well as
$\sigma_v$ and $L$. The three quantities ($R$, $L$, $\sigma_v$) are
linked together in the relation
\begin{displaymath}
(L/L_{\odot}) \propto (R/kpc)^{\alpha} \times (\sigma_v/km/s)^{\beta}
\end{displaymath}
which is the FP of galaxy clusters. An edge-on view of this FP is
shown in Figure~\ref{f-fpkh}; a 3D representation of the same FP is
given in Figure~\ref{f-fp3d}. It is remarkable that the galaxy density
profile is best represented by a King model, and also the relation
among $R$, $L$ and $\sigma_v$ has the lowest scatter when $R$ is taken
to be the King core-radius. In this case, $\alpha=1.19 \pm 0.14$ and
$\beta=0.91 \pm 0.16$.

The comparison of our FP with that derived by Schaeffer et al. is not
straightforward. Our best-fit FP was obtained using the King
core-radius, while they had to rely on the de~Vaucouleurs effective
radii which were obtained by West, Oemler \& Dekel (1989). Had we
adopted the de~Vaucouleurs effective radii as they did, the difference
between our FP and theirs, would decrease. Given the uncertainties,
Schaeffer et al.'s FP and ours, are consistent.

The FP of our sample of clusters, {\em is} significantly different
from the FP of elliptical galaxies, mainly because of the flatter
relation between $L$ and $\sigma_v$ ($\beta = 0.91 \pm 0.16$
for clusters, and $\beta = 1.82 \pm 0.14$ for
ellipticals, see, e.g., Pahre et al. 1995). This is not surprising,
given the different physical nature of clusters and
ellipticals. Therefore, we do {\em not} agree with Schaeffer et al.'s
claim of a unique FP for globular clusters, galaxies and rich
clusters.

Our FP also differs from the virial expectation for a constant
mass-to-light ratio. It does not allow a simple power-law relation
between cluster mass and cluster light, either. The deviation from the
virial relation could either indicate a (roughly) linear increase of
$M/L$ with $\sigma_v$, or that clusters are not in virial equilibrium.
However, if clusters are not dynamically relaxed systems, it is far
from clear why they define a FP at all. An increase of $M/L$ with
$\sigma_v$, and therefore with the cluster mass, has profound
implications on the determination of the cluster mass function, and on
the generalisation of results obtained from gravitational lensing,
which generally concern the most massive clusters only.

A significant part of the scatter around the FP is intrinsic. Part of
it may be due to the incomplete virialization of some of our clusters,
as evidenced by the fact that the lower-contrast and less regular
clusters are poor tracers of the FP. Part of the scatter may come from
distance errors induced by peculiar velocities. Assuming that all the
intrinsic scatter in the FP is due to deviations from the pure Hubble
flow, we found an upper limit of $\simeq 1000$~km/s to the peculiar
velocities of our clusters (in agreement with independent estimates
by, e.g., Bahcall \& Oh 1996).

\section{The phase-space distributions of cluster galaxies}
Cluster galaxies occupy different positions in the phase-space
according to their morphologies and luminosities. Early-type galaxies
are located in denser environments and have smaller velocity
dispersion than late-types (Dressler 1980a, 
Whitmore \& Gilmore 1991, Stein 1997, and
references therein). The most luminous cluster galaxies are found in
the cluster cores, and have a lower $\sigma_v$ (Biviano et al. 1992,
and references therein).

Here we summarize recent results we have obtained on these topics,
using the ENACS data-base in combination with data from the
literature. Different cluster galaxy populations have been chosen
according to their spectral types, their morphologies, and their
luminosities. The results on phase-space segregation have been
interpreted on the basis of simple kinematical models. 

\subsection{The Different Populations of Cluster Galaxies}
We defined the following cluster galaxy populations: 
\begin{enumerate}
\item galaxies with (or without) emission-lines in their spectra
(Biviano et al. 1997);
\item galaxies with spectra classified 'early' or 'late', according to
a Principal Component/Artificial Neural Network analysis (de Theije \&
Katgert 1998);
\item galaxies of different morphological types taken from the literature
(mostly from Dressler 1980b; see Adami et al. 1998b).
\end{enumerate}

The first sample contains 3729 ENACS galaxies in 75 clusters, 559 of
which are emission-line galaxies (ELG). Since the sample is not
complete in terms of equivalent width, there may be undetected ELG
among the 3170 non-ELG.  As a matter of fact, using the ENACS data
with morphological types from Dressler (1980b), we found that almost
all cluster ELG are Spirals (S) or Irregulars (I), yet only one third
of all S and I are ELG. It is important to understand whether the
non-ELG late-types share the same phase-space distribution as the ELG
or not. Sample n.2 was built with this purpose; ENACS spectra were
classified into 1571 early-types and 1023 late-types. 

Finally, sample n.3 was built with the purpose of a finer
morphological sampling. Most of the 1998 cluster galaxy data in this
sample are taken from the literature (all morphological types and
80~\% of velocities).

\subsection{ELG (Emission-Line Galaxies) vs. non-ELG}
\begin{figure}
\plottwo{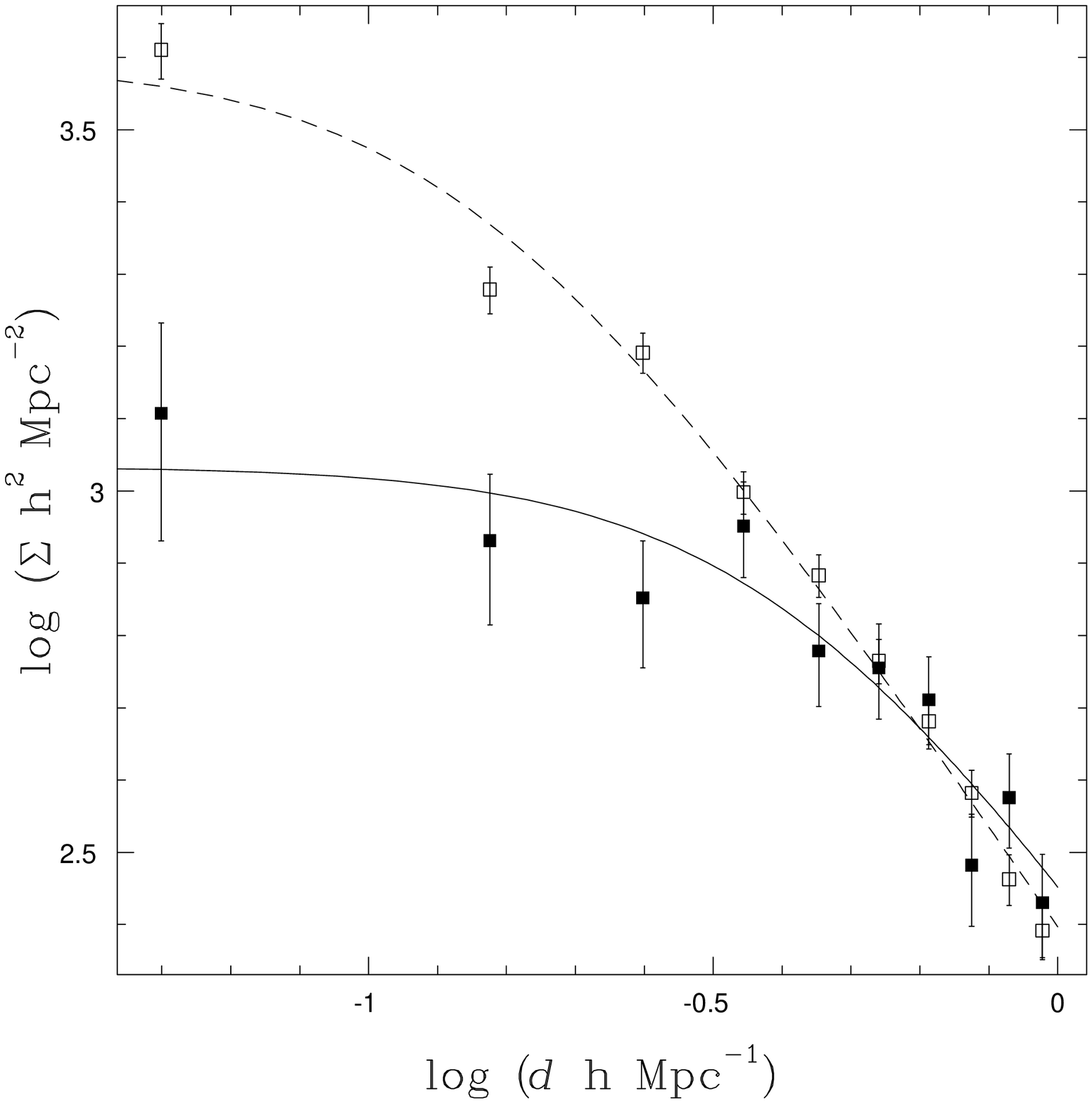}{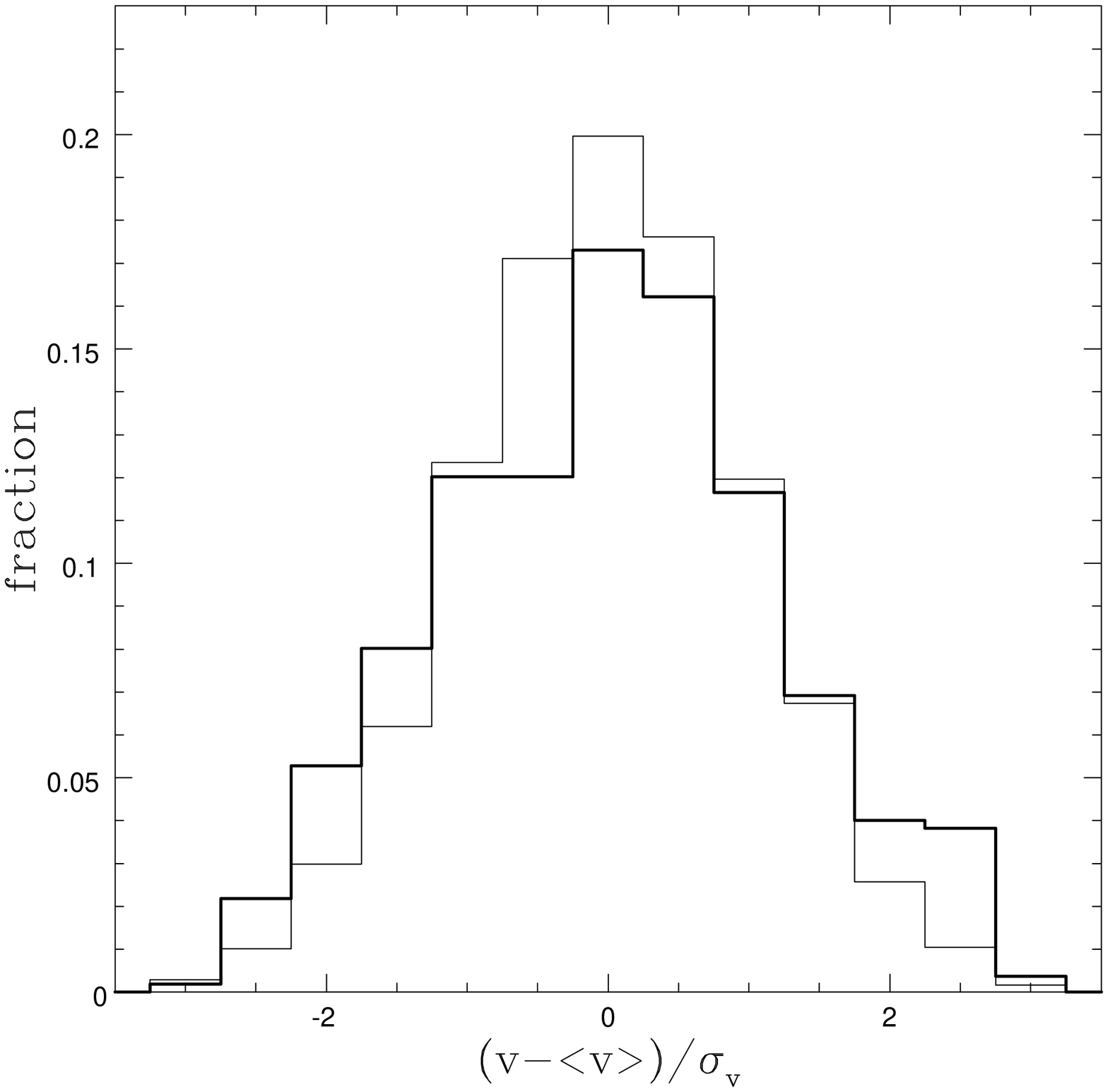}
\caption{Left panel: The number density profiles of ELG (filled
circles) and non-ELG (open circles) in a cumulative sample of 75 clusters.
The solid and dashed lines represent the best-fit beta-models to the
ELG and, respectively, the non-ELG projected density profiles.
Right panel: The normalised line-of-sight velocity distribution of ELG (thick line)
and non-ELG (solid line) in the same sample of 75 clusters.}
\label{f-dpvelg}
\end{figure}

The density profile of ELG is clearly flatter than that of non-ELG, as
can be seen in Figure~\ref{f-dpvelg} (left panel). Fitting a beta-model to the two
profiles, we estimated the ELG core radius to be three times larger
than the non-ELG core radius. Not only were the ELG found to have a
wider spatial distribution than non-ELG; they also have a wider
velocity distribution. This is illustrated in Figure~\ref{f-dpvelg} (right panel),
where the 75 cluster samples have been combined by using normalised
galaxy line-of-sight velocities, $v_n \equiv (v-<v>)/\sigma_v$. The
ELG $\sigma_v$ is $\sim 20$~\% larger than the non-ELG $\sigma_v$.

The wider spatial and velocity distributions of ELG, combine to give
an ELG-based virial mass estimate 50~\% higher than the one obtained
using the non-ELG sample. If non-ELG trace the cluster potential, ELG
are not in virial equilibrium within the same potential. Since ELG are
nonetheless cluster members\footnote{Interlopers were eliminated from
the sample using the procedure of den~Hartog \& Katgert (1996). Of
course, no interloper removal procedure is perfect. However, the
procedure is blind with respect to the galaxy type, so there is no
reason why the fraction of wrongly assigned cluster members should be
different for the ELG and the non-ELG population. Therefore
interlopers cannot explain the different virial mass estimates that
result from using one or another population.}, they must be an
unrelaxed bound cluster population. ELG have probably never crossed
the cluster core, or their emission-line properties would have been
affected by environmental related processes, while they are
undistinguishable from those of field ELG (Biviano et al. 1997).

The velocity dispersion profile (vdp, hereafter) is a useful
(contracted) representation of the phase-space distribution.  The
non-ELG vdp is well fitted by a kinematical model with zero anisotropy
and a decreasing radial velocity dispersion profile, that extrapolates
to zero at a clustercentric distance of 8~h$^{-1}$~Mpc (close to a
typical cluster turn-around radius). The ELG vdp, on the other hand,
is too steep to be fitted with a zero-anisotropy model. A good fit is
obtained with a model with constant anisotropy:
\begin{displaymath}
{\cal A} \equiv 1-\sigma_t^2/\sigma_r^2 \simeq 0.5
\end{displaymath}
where $\sigma_t$ and $\sigma_r$ are the tangential and, respectively,
the radial component of the velocity dispersion. Based on the ELG vdp
and on their gas content, we are led to the conclusion that ELG are
on their first infall onto the cluster.

\subsection{Early- vs. Late-Spectral Type Galaxies}
The PCA/ANN method was applied to the ENACS spectra to discriminate
among early- and late-types (de Theije \& Katgert 1998). Moreover, the
late-type galaxy sample contained enough ELG that it was possible to
investigate separately the phase-space distributions of the late-type
ELG and the late-type non-ELG.

The results are: 
\begin{itemize}
\item the density profile of the late-type galaxies is flatter than
that of the early-type galaxies, but the density profile of the
late-type ELG is even flatter (see Figure~4 in Katgert et al., these
proceedings);
\item the late-type galaxy $\sigma_v$ is 12~\% larger than the early-type
galaxy $\sigma_v$. However, the $\sigma_v$ of the non-ELG late-type galaxies
is almost identical to that of the early-types, while the late-type ELG
$\sigma_v$  is 25~\% higher than that of the early-types.
\end{itemize}

Therefore, the phase-space distribution of the non-ELG late-type
galaxies, is closer to that of the early-type galaxies, than to that
of the ELG late-type galaxies. It seems that only the actively star
forming galaxies do have a quite different phase-space distribution
from other cluster galaxies. Once a late-type galaxy is stripped off
its gas, it also reduces its velocity, or its orbital radial
anisotropy (or both). We must then look for an environment-related
physical process which can, at the same time, affect the galaxy gas
content, and its orbit.

\subsection{The Hubble Sequence}
Since the PCA/ANN method only allowed us to distinguish two
morphological classes, we complemented the ENACS data-set with data
from the literature, in order to build a sample of $\sim 2000$
galaxies with velocities, and morphological types. We analysed the
phase-space distribution of four different morphological classes:
ellipticals (E), S0's, early spirals (Se $\equiv$ Sa--Sbc), and late
spirals (Sl $\equiv$ Sc--Sm $+$ irregulars).

\begin{figure}
\centerline{\psfig{file=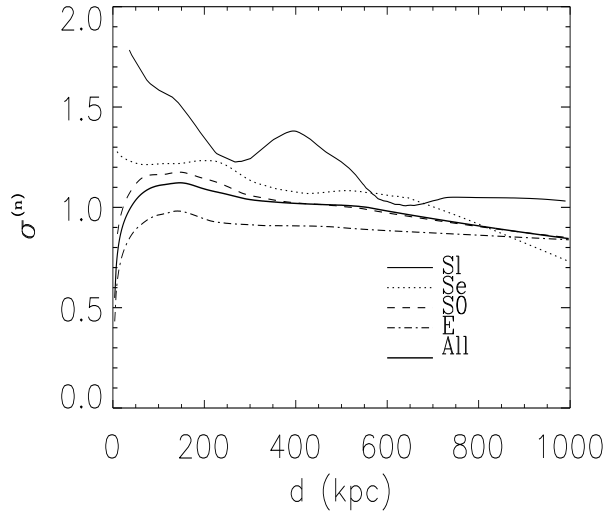,height=8cm,width=9cm}}
\caption{The velocity dispersion profile of E, S0, Se, and Sl, and of
the four classes together.}
\label{f-vdpmorph}
\end{figure}

Fitting the density profiles of the four classes with a King function,
we found a monotonic increase of the core-radius along the Hubble
sequence. The $\sigma_v$ of E is 13~\% lower than that of S0, which is
similar to that of Se. The Sl $\sigma_v$ is 20~\% higher than that of
Se. The vdp's of the four classes are also different (see
Figure~\ref{f-vdpmorph}). First, Se and Sl have a decreasing vdp in
the centre, while E and S0 have an increasing vdp out to 0.2~h$^{-1}$
Mpc. Second, the vdp's of E, S0 and Se are similar for clustercentric
distances larger than 0.2~h$^{-1}$ Mpc, while the Sl vdp is steeper
and (almost) always above all the others.

The E and S0 vdp's can be fitted using simple kinematical models with
significant tangential anisotropy in the core, and isotropy outside.
The Se and Sl vdp's, instead, require radial anisotropy, higher for
the Sl (${\cal A} \simeq 0.6$). The low $\sigma_v$ of E (and, to a
less extent, S0) in the core, is an evidence for luminosity
segregation, since it concerns the brightest cluster galaxies
only. Luminosity segregation is indeed visible in our data (see
Figure~1a in Adami et al. 1998b), and probaly results from merging
and/or dynamical friction, as predicted by theoretical models (Menci
\& Fusco-Femiano 1996).  On the other hand, the large $\sigma_v$ of Sl
suggests that these galaxies form a bound yet unvirialized population,
possibly on a first infall onto the cluster. The same conclusion was
reached by Biviano et al. (1997) about the ELG. Our results are
instead in total disagreement with Ram\'{\i}rez \& de Souza (1998).

Unfortunately, since a large part of this sample is drawn from the
literature, spectra are unavailable for most of these galaxies, and we
do not have any information on their emission-line properties.  The
similarity we see between the phase-space distribution of Sl and ELG
could well be a tautology (Sl$=$ELG), but the present sample cannot
tell us whether the fraction of ELG among Sl is significantly larger
than the fraction of ELG among other classes, Se included. 

\section{Conclusions}
We have reviewed the main results obtained by using
the ENACS data-set of cluster galaxy velocities. The main conclusions
we have reached are the followings:
\begin{enumerate}
\item{The $\sigma_v$ distribution}
\begin{itemize}
\item the intra-cluster gas 
and the cluster galaxies are in equilibrium in the same gravitational potential;
\item in standard CDM models, a bias $\geq 2$ is required.
\end{itemize}
\item{The FP of galaxy clusters}
\begin{itemize}
\item it exists, and it is different from the FP of ellipticals, and
also from virial predictions with constant $M/L$;
\item if the tilt of the FP is due to different $M/L$ for different clusters,
$M/L \propto \sigma_v$;
\item less regular clusters are also less dynamically relaxed;
\item peculiar velocities of clusters must be $\leq 1000$~km/s.
\end{itemize}
\item{The phase-space distributions of cluster galaxies}
\begin{itemize}
\item ELG and late-type spirals are bound but probably non-fully
virialized cluster populations, maybe on first infall onto the
cluster;
\item there is evidence for luminosity segregation of the brightest ellipticals
and S0's, which suggest that these galaxies have undergone significant
dynamical evolution (via the dynamical friction and merging processes).
\end{itemize}
\end{enumerate}

\acknowledgments

We thank ESO for the allocation of the observing time. We acknowledge
contributions of other members of the ESO Cluster Key Programme. We
would like to thank the workshop organizers for a very enjoyable meeting.

\end{document}